# Detecting Security threats in the Router using Computational Intelligence

J.Visumathi
Research Scholar
Sathyabama University,
Chennai-600 119
jsvisu@gmail.com

Dr. K. L. Shunmuganathan
Professor & Head, Department of CSE
R.M.K. Engineering College
Chennai-601 206
Kls_nathan@yahoo.com

**Abstract**

Information security is an issue of global concern. As the Internet is delivering great convenience and benefits to the modern society, the rapidly increasing connectivity and accessibility to the Internet is also posing a serious threat to security and privacy, to individuals, organizations, and nations alike. Finding effective ways to detect, prevent, and respond to intrusions and hacker attacks of networked computers and information systems. This paper presents a knowledge discovery frame work to detect DoS attacks at the boundary controllers (routers). The idea is to use machine learning approach to discover network features that can depict the state of the network connection. Using important network data (DoS relevant features), we have developed kernel machine based and soft computing detection mechanisms that achieve high detection accuracies. We also present our work of identifying DoS pertinent features and evaluating the applicability of these features in detecting novel DoS attacks. Architecture for detecting DoS attacks at the router is presented. We demonstrate that highly efficient and accurate signature based classifiers can be constructed by using important network features and machine learning techniques to detect DoS attacks at the boundary controllers.

## 1   Introduction

By nature Internet is public, connected, distributed, open, and dynamic. Phenomenal growth of computing devices, connectivity speed, and number of applications running on networked systems posed engendering risk to the Internet. Malicious usage, attacks, and sabotage have been on the rise as more and more computing devices are put into use. Connecting information systems to networks such as the Internet and public telephone systems further magnifies the potential for exposure through a variety of attack channels. These attacks take advantage of the flaws or omissions that exist within the various information systems and software that run on many hosts in the network.

In DoS attacks the adversary mainly targets a few services like network bandwidth, router or server CPU cycles, system storage, operating system data structures, protocol data structures and software vulnerabilities. DoS can be a single source attack, originating at a single host, or can be a multi-source source attack, where multiple hosts and networks are involved. The DoS attacks can take an advantage form the distributed nature of the Internet by launching a multiplicative effect, resulting in distributed DoS. Due to the use of dynamic protocols and address spoofing, detecting distributed and automated attacks still remains a challenge.

Efforts on how to define and characterize denial of service attacks through a collection of different perspectives such as bandwidth, process information, system information, user information and IP address is being proposed by several researchers [1,6]. Using the defined characteristics a few signature-based and anomaly based detection techniques are proposed [2,9]. Recent malware and distributed DoS attacks proved that there exists no effective means to detect, respond and mitigate availability attacks.

In this paper we propose a router based approach to detect denial of service attacks using intelligent systems. A comparative study of support vector machines (SVMs), Multi adaptive regression splines (MARSs) and linear genetic programs (LGPs) for detecting denial of service attacks is performed through a variety of experiments performed on a well know Lincoln Labs data set that consists of more than 80% of different denial of service attacks described in section 2. We address the use of machine learning approach to discover network features that can depict the state of the network connection. We also present our work of identifying DoS pertinent features from a publicly available intrusion detection data set and evaluating the applicability of these features in detecting novel DoS attacks on a live performance network. Architecture for detecting DoS attacks at the routers.

In the rest of the paper, a brief introduction to the data used and DoS attacks is given in section 2. An overview of soft computing paradigms used is given in section 3. Experiments for detecting DoS attacks using MARs, SVMs and LGPs are given in section 4. Significant feature identification techniques are presented in section 5. In section 6 we present the architecture and the applicability of DoS significant features in detecting DoS attacks at the routers. Conclusions are presented in section 7.

## 2 Intrusion detection data

A sub set of the DARPA intrusion detection data set is used for off-line analysis. In the DARPA intrusion detection evaluation program, an environment was set up to acquire raw TCP/IP dump data for a network by simulating a typical U.S. Air Force LAN. The LAN was operated like a real environment, but being blasted with multiple attacks [5,11]. For each TCP/IP connection, 41 various quantitative and qualitative features were extracted [16]. The 41 features extracted fall into three categories, "intrinsic" features that describe about the individual TCP/IP connections; can be obtained form network audit trails, "content-based" features that describe about payload of the network packet; can be obtained from the data portion of the network packet, "traffic-based" features, that are computed using a specific window.

### 2.1 Denial of service attacks

Attacks designed to make a host or network incapable of providing normal services are known as denial of service attacks. There are different types of DoS attacks: a few of them abuse the computers legitimate features; a few target the implementations bugs; and a few exploit the misconfigurations. DoS attacks are classified based on the services that an adversary makes unavailable to legitimate users. A few examples include preventing legitimate network traffic, preventing access to services for a group or individuals. DoS attacks used for offline experiments and identifying significant features are presented in table 1 [5,11].

**TABLE 1: DoS Attack Description**

| Attack Type | Service | Effect of the attack |
|---|---|---|
| Apache2 | http | Crashes httpd |
| Land | http | Freezes the machine |
| Mail bomb | N/A | Annoyance |
| SYN Flood | TCP | Denies service on one or more ports |
| Ping of Death | Icmp | None |
| Process table | TCP | Denies new processes |
| Smurf | Icmp | Slows down the network |
| Syslogd | Syslog | Kills the Syslogd |
| Teardrop | N/A | Reboots the machine |
| Udpstrom | Echo | Slows down the network |

## 3 Soft computing paradigms

Soft computing was first proposed by Zadeh to construct new generation computationally intelligent hybrid systems consisting of neural networks, fuzzy inference system, approximate reasoning and derivative free optimization techniques. It is well known that the intelligent systems, which can provide human like expertise such as domain knowledge, uncertain reasoning, and adaptation to a noisy and time varying environment, are important in tackling practical computing problems. In contrast with conventional Artificial Intelligence (AI) techniques which only deal with precision, certainty and rigor the guiding principle of hybrid systems is to exploit the tolerance for imprecision, uncertainty, low solution cost, robustness, partial truth to achieve tractability, and better rapport with reality

### 3.1 Support vector machines

The SVM approach transforms data into a feature space F that usually has a huge dimension. It is interesting to note that SVM generalization depends on the geometrical characteristics of the training data, not on the dimensions of the input space [3,4]. Training a support vector machine (SVM) leads to a quadratic optimization problem with bound constraints and one linear equality constraint. Vapnik shows how training a SVM for the pattern recognition problem leads to the following quadratic optimization problem .
Minimize:

$$W(\alpha) = -\sum_{i=1}^{l} \alpha_i + \frac{1}{2} \sum_{i=1}^{l} \sum_{j=1}^{l} y_i y_j \alpha_i \alpha_j k(x_i, x_j) \quad (1)$$

Subject to $\sum_{i=1}^{l} y_i \alpha_i$  (2)

$$\forall i : 0 \leq \alpha_i \leq C$$

Where l is the number of training examples $\alpha$ is a vector of l variables and each component $\alpha_i$ corresponds to a training example ($x_i$, $y_i$). The solution of (1) is the vector $\alpha^*$ for which (1) is minimized and (2) is fulfilled.

### 3.2 Linear genetic programs

LGP is a variant of the Genetic Programming (GP) technique that acts on linear genomes . The linear genetic programming technique used for our current experiment is based on machine code level manipulation and evaluation of programs. Its main characteristics in comparison to tree-based GP lies is that the evolvable units are not the expressions of a functional programming language (like LISP), but the programs of an imperative language (like C) are evolved. In the Automatic Induction of Machine Code by Genetic Programming, individuals are manipulated directly as binary code in memory and executed directly without passing an interpreter during fitness calculation. The LGP tournament selection procedure puts the lowest selection pressure on the individuals by allowing only two individuals to participate in a tournament. A copy of the winner replaces the loser of each tournament. The crossover points only occur between instructions. Inside instructions the mutation operation randomly replaces the instruction identifier, a variable or the constant from valid ranges. In LGP the maximum size of the program is usually restricted to prevent programs without bounds. As LGP could be implemented at machine code level, it will be fast to detect intrusions in a near real time mode.

### 3.3 Multi adaptive regression splines

Splines can be considered as an innovative mathematical process for complicated curve drawings and function approximation. To develop a spline the X-axis is broken into a convenient number of regions. The boundary between regions is also known as a knot. With a sufficiently large number of knots virtually any shape can be well approximated. While it is easy to draw a spline in 2-dimensions by keying on knot locations (approximating using linear, quadratic or cubic polynomial etc.), manipulating the mathematics in higher dimensions is best accomplished using basis functions. The MARS model is a regression model using basis functions as predictors in place of the original data. The basis function transform makes it possible to selectively blank out certain regions of a variable by making them zero, and allows MARS to focus on specific sub-regions of the data. It excels at finding optimal variable transformations and interactions, and the complex data structure that often hides in high-dimensional data .

### 4 Offline evaluation

We partition the data into the two classes of "Normal" and "DoS" patterns, where the DoS attack is a collection of six different attacks (back, neptune, ping of death, land, smurf, and teardrop). The objective is to separate normal and DoS patterns. The (training and testing) data set contains 11982 randomly generated from data described in section 3, with the number of data from each class proportional to its size, except that the smallest class is completely included. A different randomly selected set of 6890 points of the total data set (11982) is used for testing different soft computing paradigms. Results of SVM, MARS and LGP classifications are given in Table 2.

**TABLE 2: Classifier Evaluation for Offline DoS Data**

| Class | Classifier Accuracy (%) | | |
|---|---|---|---|
| | SVM | LGP | MARS |
| Normal | 98.42 | 99.64 | 99.71 |
| DoS | 99.45 | 99.90 | 96 |

### 5 Significant feature identification

Feature selection and ranking is an important issue in intrusion detection. Of the large number of features that can be monitored for intrusion detection purpose, which are truly useful, which are less significant, and which may be useless? The question is relevant because the elimination of useless features enhances the accuracy of detection while speeding up the computation, thus improving the overall performance of an IDS. In cases where there are no useless features, by concentrating on the most important ones we may well improve the time performance of an IDS without affecting the accuracy of detection in statistically significant ways.

- Having a large number of input variables $x = (x_1, x_2, \ldots, x_n)$ of varying degrees of importance to the output y; i.e., some elements of x are essential, some are less important, some of them may not be mutually independent, and some may be useless or irrelevant (in determining the value of y)

- Lacking an analytical model that provides the basis for a mathematical formula that precisely describes the input-output relationship, $y = F(x)$

- Having available a finite set of experimental data, based on which a model (e.g. neural networks) can be built for simulation and prediction purposes

### 5.1 Support vector decision function ranking

Information about the features and their contribution towards classification is hidden in the support vector decision function. Using this information one can rank their significance, i.e., in the equation

$$F(X) = \Sigma W_i X_i + b \quad (3)$$

The point X belongs to the positive class if F(X) is a positive value. The point X belongs to the negative class if F(X) is negative. The value of F(X) depends on the contribution of each value of X and $W_i$. The absolute value of $W_i$ measures the strength of the classification. If $W_i$ is a large positive value then the $i^{th}$ feature is a key factor for positive class. If $W_i$ is a large negative value then the $i^{th}$ feature is a key factor for negative class. If $W_i$ is a value close to zero on either the positive or the negative side, then the $i^{th}$ feature does not contribute significantly to the classification. Based on this idea, a ranking can be done by considering the support vector decision function.

*5.2    Linear genetic ranking algorithm*

In the feature selection problem the interest is in the representation of the space of all possible subsets of the given input set. An individual of length d corresponds to a d-dimensional binary feature vector Y, where each bit represents the elimination or inclusion of the associated feature. Then, $y_i = 0$ represents elimination and $y_i = 1$ indicates inclusion of the $i^{th}$ feature. Fitness F of an individual program p is calculated as the mean square error (MSE) between the predicted output ($O_{ij}^{pred}$) and the desired output ($O_{ij}^{des}$) for all n training samples and m outputs.

$$F(p) = \frac{1}{n \cdot m} \sum_{i=1}^{n} \sum_{j=1}^{m} (O_{ij}^{pred} - O_{ij}^{des})^2 + \frac{w}{n} CE = MSE + w \cdot MCE \quad (4)$$

Classification Error (*CE*) is computed as the number of misclassifications. Mean Classification Error (*MCE*) is added to the fitness function while its contribution is proscribed by an absolute value of Weight (W).

*5.3    Multi adaptive regression spines ranking*

Generalized cross-validation is an estimate of the actual cross-validation which involves more computationally intensive goodness of fit measures. Along with the MARS procedure, a generalized cross-validation procedure is used to determine the significant input features. Non-contributing input variables are thereby eliminated.

$$GCV = \frac{1}{N} \sum_{i=1}^{N} [\frac{y_i - f(x_i)^2}{1 - k/N}] \quad (5)$$

where N is the number of records and x and y are independent and dependent variables respectively. k is the effective number of degrees of freedom whereby the GCV adds penalty for adding more input variables to the model.

*5.4    Significant feature off line evaluation*

Description of most important features as ranked by three feature-ranking algorithms (SVDF, LGP, and MARS) is given in table 3. Classifier performance using all the 41 features and most important 6 features as inputs to the classifier is given in table 4.

**TABLE 3: Classifier Evaluation for Offline DoS Data**

| Ranking | Feature Description |
|---|---|
| SVDF | count:<br>service count:<br>dst ost rv_serror_rate<br>serror_rate:<br>dst_host_same_src_port_rate<br>dst_host_serror_rate: % |
| LGP | count:<br>compromised conditions:<br>wrong fragments:<br>land:<br>logged in:<br>hot: |
| MARS | count:<br>service count:<br>dst_host_srv_diff_host_rate:<br>source bytes:<br>destination bytes:<br>hot: |

**TABLE 4: Significant feature evaluation**

| Classifier | Features | | | |
|---|---|---|---|---|
| | Normal 41 | DoS 41 | Normal 6 | DoS 6 |
| SVM | 98.42 | 99.45 | 99.23 | 99.16 |
| LGP | 99.64 | 99.90 | 99.77 | 99.14 |
| MARS | 99.71 | 96 | 99.80 | 95.47 |

**6    Real time router based DoS detection**

A passive sniffer can be placed at the router to collect data for detecting DoS attacks.

The architecture comprises of three components: a packet parser, classifier and a response module. The network packet parser uses the WINPCAP library to capture packets and extracts the relevant features required for DoS detection. The output of the parser includes the twelve DoS-relevant features as selected by our ranking algorithm [7,8].

The output summary of the parser includes the eleven features of duration[1] of the connection to the target machine, protocol[2] used to connect, service type[3], status of the connection[4] (normal or error), number of source bytes[5], number of destination bytes[6], number of connections to the same host as the current one during a specified time window[7] (in our case .01seconds), number of connections to the same host as the current one using same service during the past 0.01 seconds[8], percentage of connections that have SYN errors during the past .01 seconds[9], percentage of connections that have SYN errors while using the same service during the past .01 seconds[10], and percentage of connections to the same service during the past .01 seconds[11].

We experimented with more than 24 types of DoS attacks, including 6 types of DoS described in section 3 and 17 additional types. In the experiments performed we used different types of DoS attacks: SYN flood, SYN full, MISFRAG, SYNK, Orgasm, IGMP flood, UDP flood, Teardrop, Jolt, Overdrop, ICMP flood, FIN flood, and Wingate crash, with different service and port options. Normal data included multiple sessions of http, ftp, telnet, SSH, http, SMTP, pop3 and imap. Network data originating from a host to the server that included both normal and DoS is collected for analysis; for proper labeling of data for training the classifier normal data and DoS data are collected at different times.

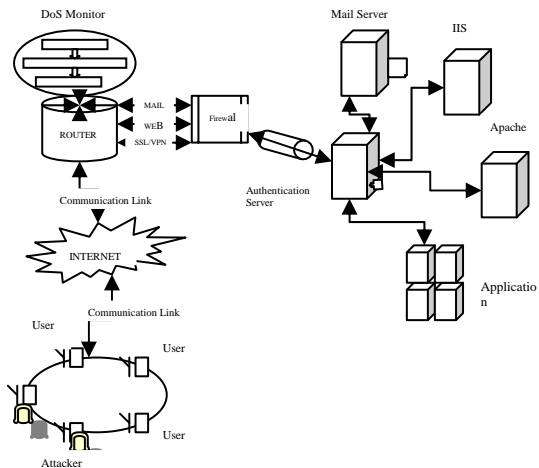

Figure 2. Architecture for detecting DoS attacks at the routers

**TABLE 5: Router based detection accuracies**

| Class/ Learning Machine | Normal SVM LGP MARS | DoS SVM LGP MARS | Accuracy SVM LGP MARS |
|---|---|---|---|
| Normal | 2692 2578 1730 | 14 128 976 | 99.48 95.26 63.9 |
| DoS | 538 153 0 | 2141 2526 2679 | 79.91 94.28 100 |
| Accuracy SVM LGP MARS | 83.77 99.08 63.9 | 80.44 99.06 73.2 | |

The top-left entry of Table 5 shows that 2692, 2578, and 1730 of the actual "normal" test set were detected to be normal by SVM, LGP and MARS respectively; the last column indicates that 99.46, 95.26 and 63.9 % of the actual "normal" data points were detected correctly. In the same way, for "DoS" 2141, 2526 and 2679 of the actual "attack" test set were correctly detected; the last column indicates that 79.91, 94.28 and 100 % of the actual "DoS" data points were detected correctly. The bottom row shows that 83.77, 99.08 and 63.0 % of the test set said to be "normal" indeed were "normal" and 83.77, 99.06 and 73.2 % of the test set classified, as "DoS" indeed belongs to DoS as classified by SVM, LGP and MARS respectively.

## 7  Conclusions

A number of observations and conclusions are drawn from the results reported:

- A comparison of different soft computing techniques is given. Linear genetic programming technique outperformed SVM and MARS with a 94.28 % detection rate on the real time test dataset.

Regarding significant feature identification, we observe that

- The three feature ranking methods produce largely consistent results.

- The most significant features for the two classes of 'Normal' and 'DOS' heavily overlap.

- Using the 6 important features for each class gives the remarkable performance.

Regarding real time router based DoS detection, we observe that

- DoS attacks can be detected at the router, thus pushing the detection as far out as possible in the network perimeter

- "Third generation worms" can be detected by tuning the time based features.

- "Low and slow" DoS attacks can be detected by judiciously selecting the time based and connection based features.

**Acknowledgement**

We take immense pleasure in thanking our Chairman Dr. Jeppiaar M.A, B.L, Ph.D, the Directors of Jeppiaar Engineering College Mr. Marie Wilson, B.Tech, MBA.,(Ph.D) Mrs. Regeena Wilson, B.Tech, MBA., (Ph.D) and the Principal Dr. Sushil Lal Das M.Sc(Engg.), Ph.D for their continual support and guidance. We would like to extend our thanks to my guide, our friends and family members without whose inspiration and support our efforts would not have come to true. Above all, we would like to thank God for making all our efforts success.

### References

[1]  W. J. Blackert, D. C. Furnanage, and Y. A. Koukoulas, "Analysis of Denial of service attacks Using An address Resolution Protocol Attack", Proc. of the 2002 IEEE Workshop on Information Assurance, US Military Academy, pp. 17-22, 2002.

[2]  D. W. Gresty, Q. Shi, and M. Merabti, "Requirements for a general framework for response to distributed denial of service," Proc. Of Seventeenth Annual Computer Security Applications Conference, pp. 422-229, 2001.

[3]  T. Joachims, "Making Large-Scale SVM Learning Practical", LS8-Report, University of Dortmund, LS VIII-Report, 2000.

[4]  T. Joachims, "SVMlight is an Implementation of Support Vector Machines (SVMs) in C", University of Dortmund. Collaborative *Research Center on Complexity Reduction in Multivariate Data (SFB475)*, 2000.

[5]  K. Kendall, "A Database of Computer Attacks for the Evaluation of Intrusion Detection Systems", *Master's Thesis, Massachusetts Institute of Technology*, 1998.

[6]  J. Mirkovic, J. Martin, and P. Reiher, "A Taxonomy of DDoS Attacks and DDoS Defense Mechanisms", *Technical Report # 020017*, Department of Computer Science, UCLA, 2002.

[7]  S. Mukkamala, and A. H. Sung, "Identifying Key Features for Intrusion Detection Using Neural Networks", Proc. ICCC International Conference on Computer Communications, pp. 1132-1138, 2002.

[8]  S. Mukkamala, and A.H. Sung "Feature Selection for Intrusion Detection Using Neural Networks and Support Vector Machines", *Journal of the Transportation Research Board of the National Academics*, Transportation Research Record No 1822, pp. 33-39, 2003.

[9]  C. Shields, "What do we mean by network denial of service?", Proc. of the 2002 IEEE workshop on Information Assurance. US Military Academy, pp. 196-203, 2002.

[10]  J. Stolfo, F. Wei, W. Lee, A. Prodromidis, and P. K. Chan, "Cost-based Modeling and Evaluation for Data Mining with Application to Fraud and Intrusion Detection", *Results from the JAM Project by Salvatore*, 1999.

[11]  S. E. Webster, "The Development and Analysis of Intrusion Detection Algorithms", *S.M. Thesis, Massachusetts Institute of Technology*, 1998.

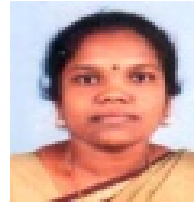

J.Visumathi B.E.,M.E.,(Ph.D) works as Assistant Professor in Jeppiaar Engineering College, Chennai and She has more than 10 years of teaching experience and her areas of specializations are Networks, Artificial Intelligence, and DBMS.

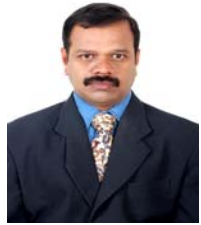

Dr. K.L. Shunmuganathan B.E, M.E.,M.S.,Ph.D works as the Professor & Head of CSE Department of RMK Engineering College, Chennai, TamilNadu, India. He has has more than18 years of teaching experience and his areas of specializations are Networks, Artificial Intelligence, and DBMS.